\documentclass{PoS}
\usepackage{subfigure}
\usepackage{ragged2e}
\title{Search for heavy BSM particles coupling to third generation quarks at CMS}

\ShortTitle{Search for heavy BSM particles coupling to third generation quarks at CMS}

\author{\speaker{Mareike Meyer} on behalf of the CMS Collaboration\\
        University of Hamburg\\
        E-mail: \email{mareike.meyer@desy.de}}

\abstract{Many models of physics beyond the Standard Model (SM) contain enhanced couplings to third generation particles. The predicted signatures include vector-like quarks and $t\bar{t}$ resonances. 
We present a review of non-SUSY based searches for new physics beyond the SM in final states with third-generation quarks using proton-proton collision data collected with the CMS detector at the CERN LHC. 
We analyze a wide range of final states, from multi-leptonic to entirely hadronic, and many results use novel analysis techniques to identify and reconstruct highly boosted final states that are created in these topologies. 
These techniques provide increased sensitivity to new high-mass particles over traditional search methods.}

\FullConference{XXV International Workshop on Deep-Inelastic Scattering and Related Subjects\\
		3-7 April 2017\\
		University of Birmingham, UK}

\begin{document}
\section{Introduction}
Due to their high mass of approximately $173\;\textrm{GeV}$~\cite{PDG} and the resultant large coupling strength to the Higgs boson, top quarks play an important role in many theories beyond the Standard Model (SM).
A lot of these Beyond Standard Model (BSM) theories predict the existence of new particles with enhanced coupling strengths to third generation SM quarks.
This article presents searches for BSM particles performed by the CMS experiment~\cite{CMS} using $pp$-collision data recorded at a center-of-mass energy of $13\;\textrm{TeV}$. 
Prior to the discussion of the analyses (see Sections \ref{TPrimeTZ} to \ref{LQbtau}), an introduction to used analysis tools is given (Section \ref{tools}).
\section{Analysis tools}
\label{tools}
\begin{figure}[b!]
  \centering
  \subfigure[]{
  \centering
    \includegraphics[width=0.45\textwidth]{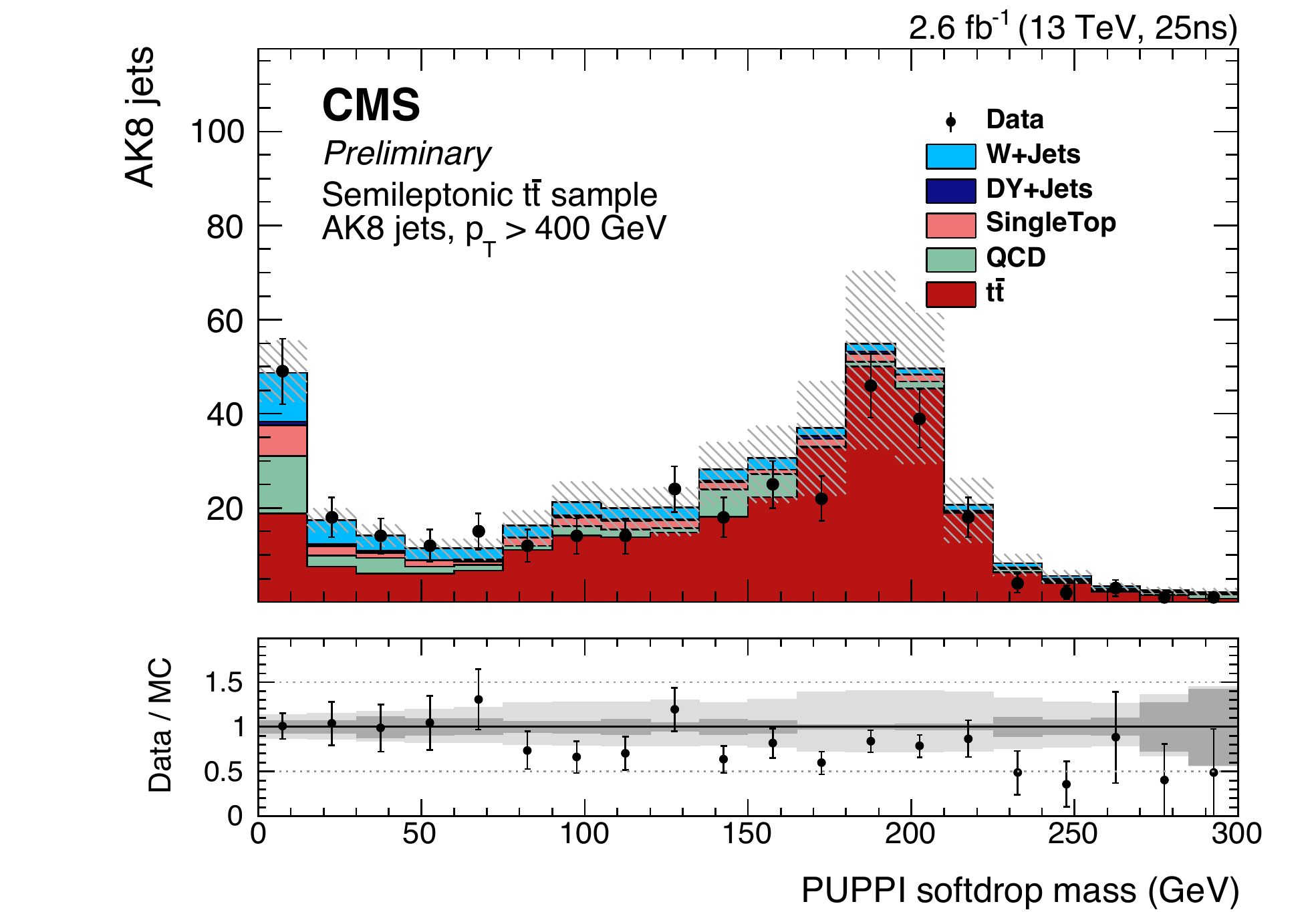}
    \label{TopTagginga}
  }
  \subfigure[]{
  \centering
    \includegraphics[width=0.45\textwidth]{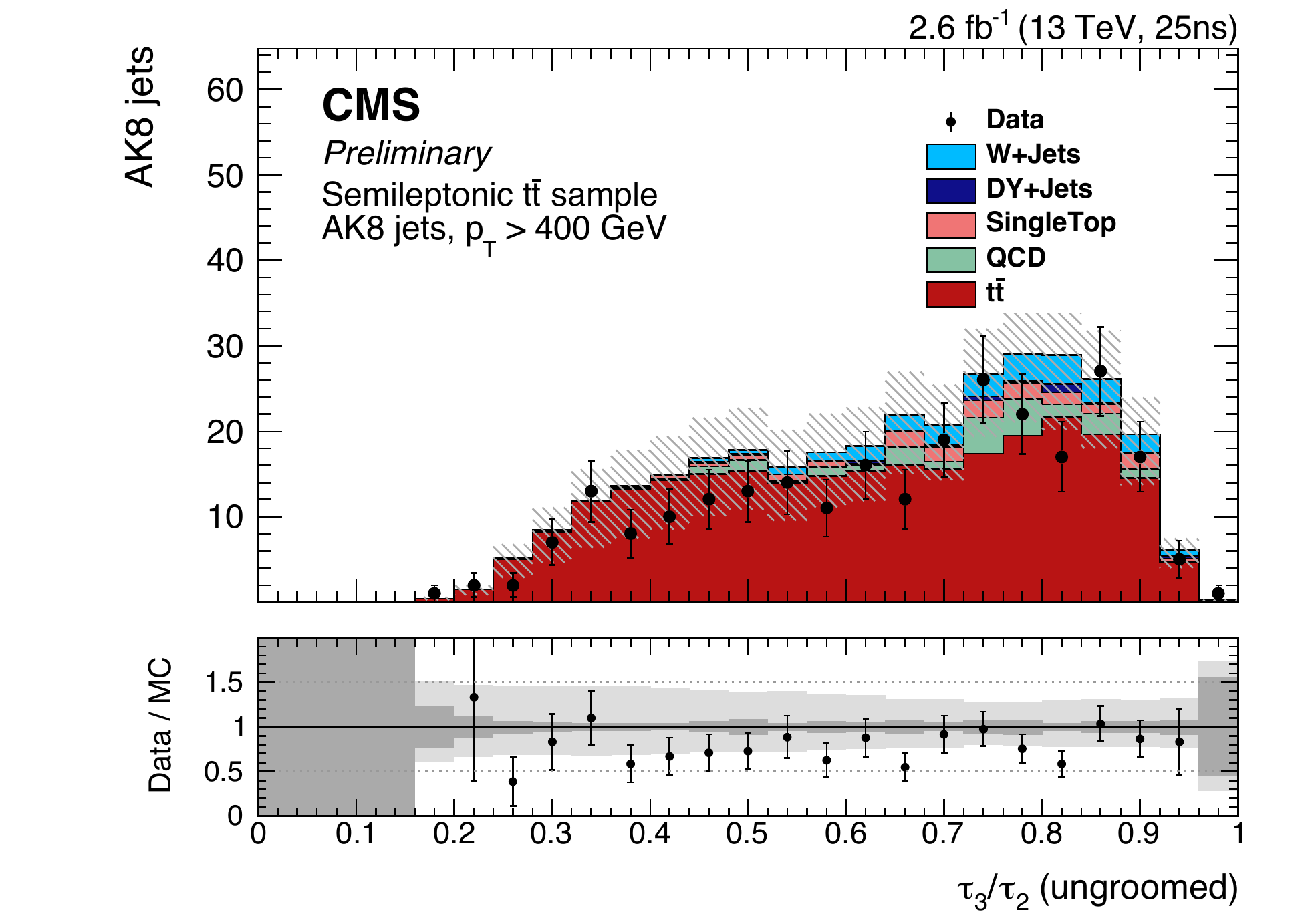}
      \label{TopTaggingb} 
  }\\
  \caption[]{Comparison of data and SM processes in the distributions of (a) the softdrop mass $m_{\textrm{SD}}$ and (b) the n-subjettiness ratio $\tau_{3}/\tau_{2}$ in a sample enriched in lepton+jets $t\bar{t}$ events. The figures are taken from~\cite{JMARPAS}.}
  \label{TopTagging}
\end{figure}
In all presented searches decays of heavy particles are studied.
Thus, the top quarks produced in the decays of these heavy objects, or at least the $W$ bosons produced in the top quark decays, receive large transverse momenta. 
Hence, their respective decay products appear very collimated.
Since traditional reconstruction techniques often fail in reconstructing the decays of boosted hadronic top quark or $W$ boson decays, the decay products are instead identified using large-cone anti-$k_T$ jets (often the distance parameter $R$ is set to $R=0.8$) together with top tagging and $W$ tagging techniques, described in~\cite{JMARPAS}.\\ 
The top tagging algorithms used in the presented analyses apply requirements on the softdrop mass $m_{\textrm{SD}}$ and on the variable n-subjettiness together with subjet $b$ tagging techniques.
In Fig.~\ref{TopTagging} the distributions for $m_{\textrm{SD}}$ and $\tau_{3}/\tau_{2}$ are presented for $t\bar{t}$ and other SM processes.
The softdrop mass $m_{\textrm{SD}}$ is the large-cone jet mass after softdrop grooming has been applied~\cite{Softdrop1,Softdrop2}.
This algorithm removes soft wide-angle radiation within a large-cone jet. 
Additional constraints on the n-subjettiness ratio $\tau_{3}/\tau_{2}$ are applied to identify the three-prong top quark decay.
The n-subjettiness variable $\tau_{i}$~\cite{NSub1, NSub2} is a measure for the compatibility with the $i$-subjet hypothesis.
To further enhance the signal over background ratio, subjet $b$ tagging, which is the application of $b$ tagging algorithms on subjets, can be applied.\\
In order to identify boosted $W$ boson decays, requirements on the softdrop mass or the pruned jet mass and the n-subjettiness ratio $\tau_{2}/\tau_{1}$ are imposed. 
Pruning is a grooming technique~\cite{Pruning}, which removes particles during the jet clustering procedure that do not pass certain distance and $p_T$ requirements.\\
To enable identification of leptonic decays of high-$p_T$ top quarks, in which standard lepton isolation requirements fail, special isolation criteria are applied. 
These include employing an isolation cone that shrinks with increasing transverse momentum of the studied lepton or applying no isolation criteria at all.
In the latter case the leptons have to pass either requirements on the distance of the lepton to the jet, or their relative momentum with respect to the jet has to exceed a certain threshold to reduce the QCD multijet background.
\section{Single production of vector-like $\mathbf{T'}$ quarks decaying to a top quark and a $\mathbf{Z}$ boson}
\label{TPrimeTZ}
\begin{figure}[bt]
  \centering
  \subfigure[]{
  \centering
    \includegraphics[width=0.4\textwidth]{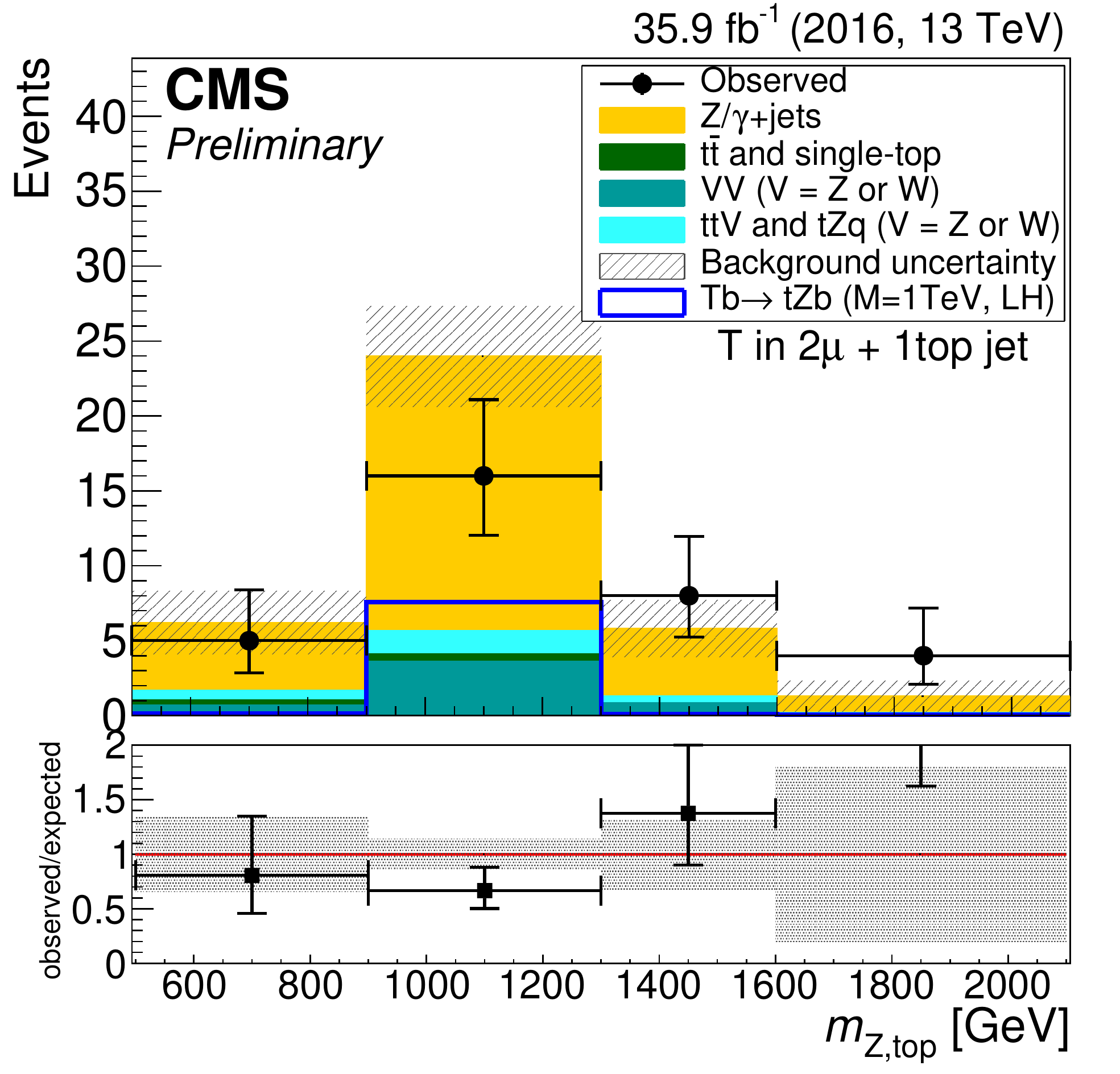}
    \label{TPrimeTZfiga}
  }
  \subfigure[]{
  \centering
    \includegraphics[width=0.4\textwidth]{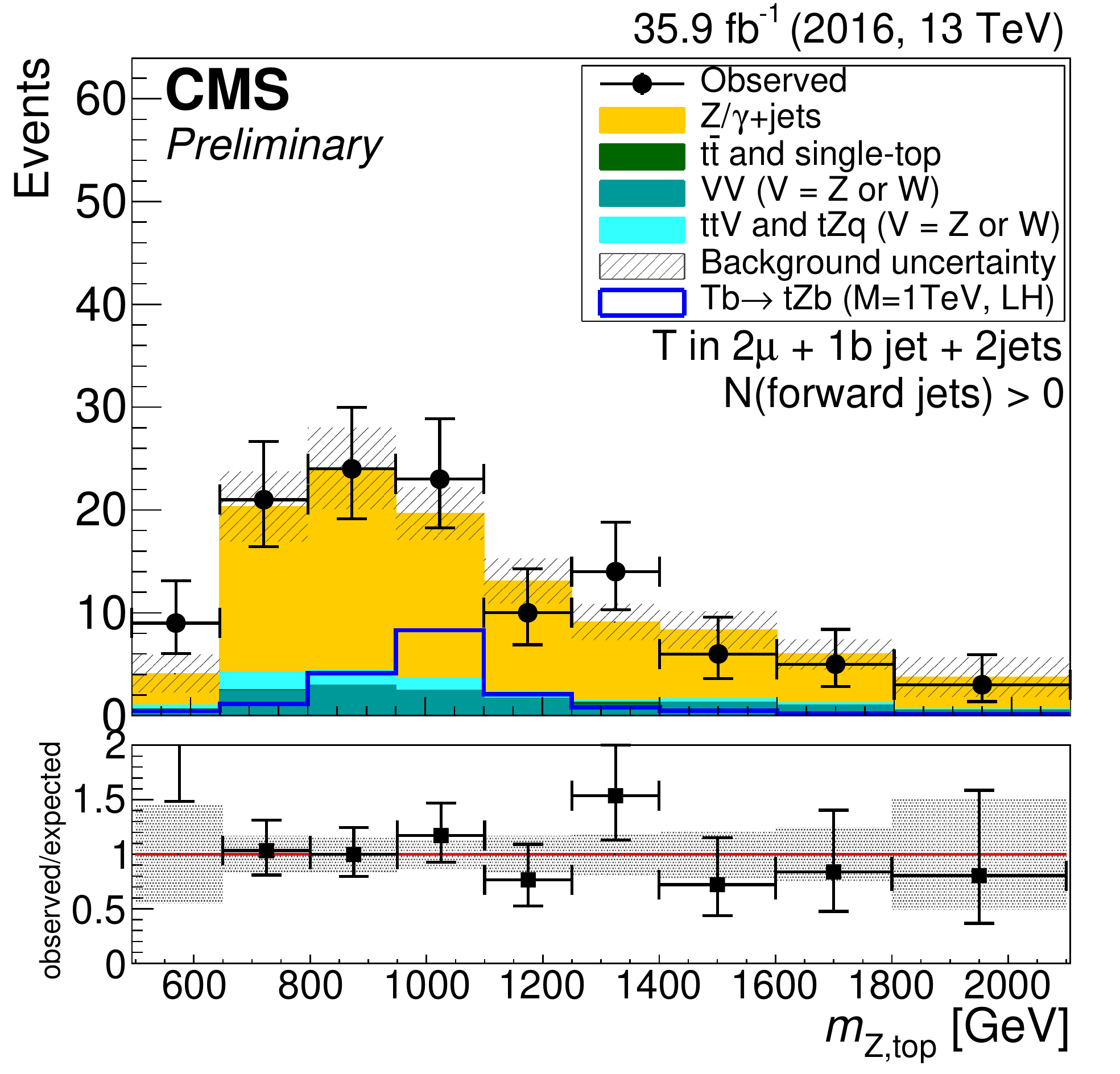}
      \label{TPrimeTZfigb} 
  }\\
  \caption[]{Comparison of data and the SM background expectation after the application of the signal selection. Presented are the distributions of the invariant mass of the reconstructed $T'$ candidate in the categories with (a) two muon candidates and a top-tagged jet and (b) with two muon candidates, a $b$-tagged jet, two additional jets and at least one forward jet. The figures are taken from~\cite{B2G17007}.}
  \label{TPrimeTZfig}
\end{figure}
A search for single production of vector-like $T'$ quarks is presented~\cite{B2G17007}. 
The analysis is based on data corresponding to $35.9\;\textrm{fb}^{-1}$ and is performed in the $T'$ decay mode to a top quark and a $Z$ boson.
The signal signature includes two leptons with an invariant mass compatible with the $Z$ boson mass, an hadronically decaying top quark candidate and a forward jet. 
The selected events are arranged in ten different categories depending on whether an electron or muon pair is reconstructed, whether or not at least one forward jet is found, and whether the hadronically decaying top quark candidate is reconstructed with a top-tagged jet, a $W$-tagged and a $b$-tagged jet, or three small-cone jets.
In the absence of a top-tagged jet, the forward jet requirement is essential for SM background suppression.
A $T'$ candidate is build for each event.
The resulting ten distributions for the invariant mass of the $T'$ candidate are used for the final statistical evaluation of the result.
Two of these distributions are presented in Fig.~\ref{TPrimeTZfig}.
In all ten event categories data agrees within statistical and systematic uncertainties with the SM expectation.
Based on the narrow-width approximation, upper exclusion limits on the single production cross section of $T'$ quarks are set in dependence of the mass of the vector-like $T'$ quark in two different models.
Most notably, this is the first search probing also models that describe $T'$ quarks with larger widths, which range between 10\% and 30\%.
For all studied widths, similar sensitivities as for the narrow-width approximation are achieved.
\section{Pair production of vector-like $\mathbf{X_{5/3}}$ quarks decaying to a top quark and a $\mathbf{W}$ boson}
\begin{figure}[tb]
  \centering
  \subfigure[]{
  \centering
    \includegraphics[width=0.4\textwidth]{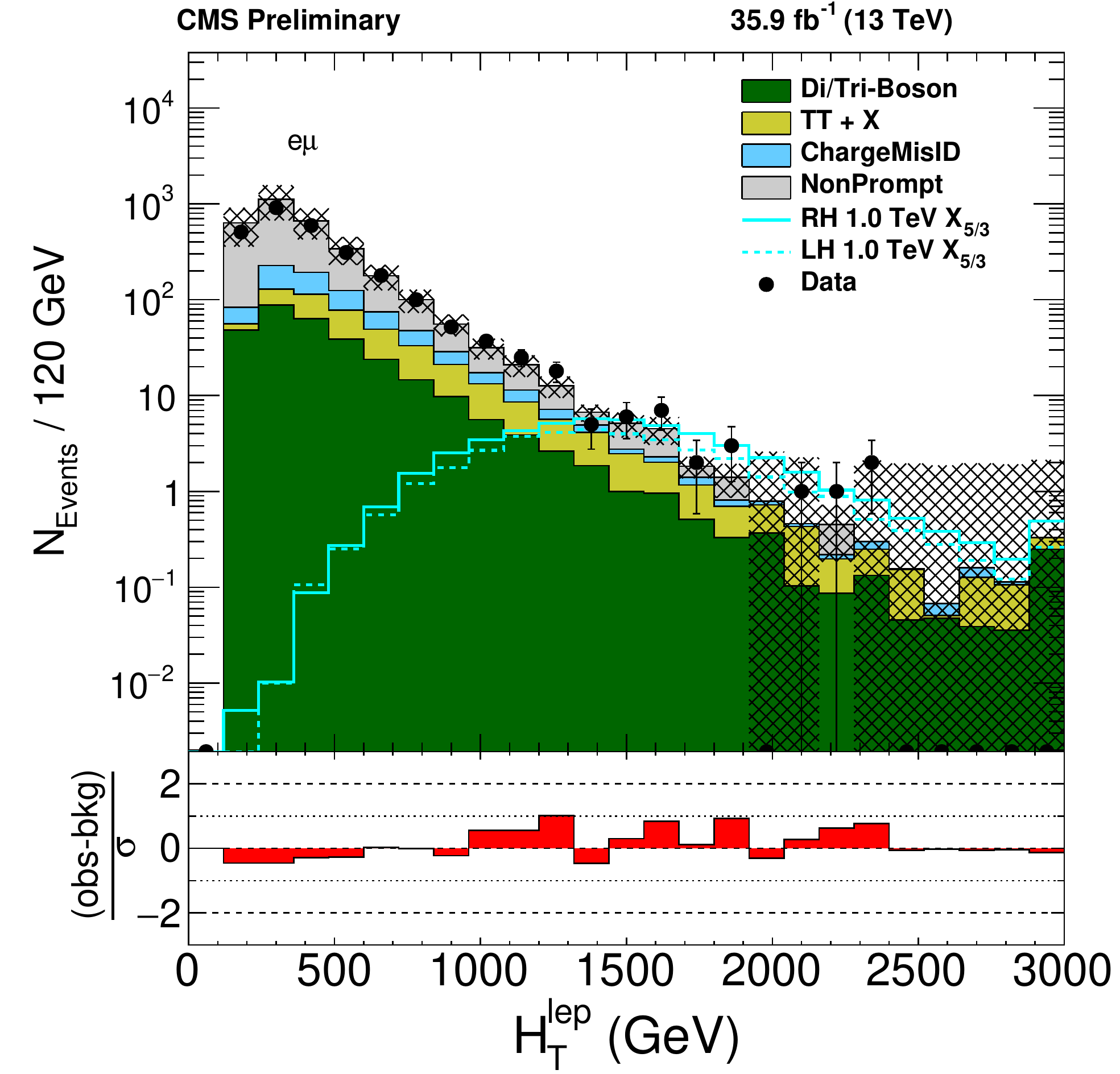}
    \label{XfiveThirdfiga}
  }
  \subfigure[]{
  \centering
    \includegraphics[width=0.4\textwidth]{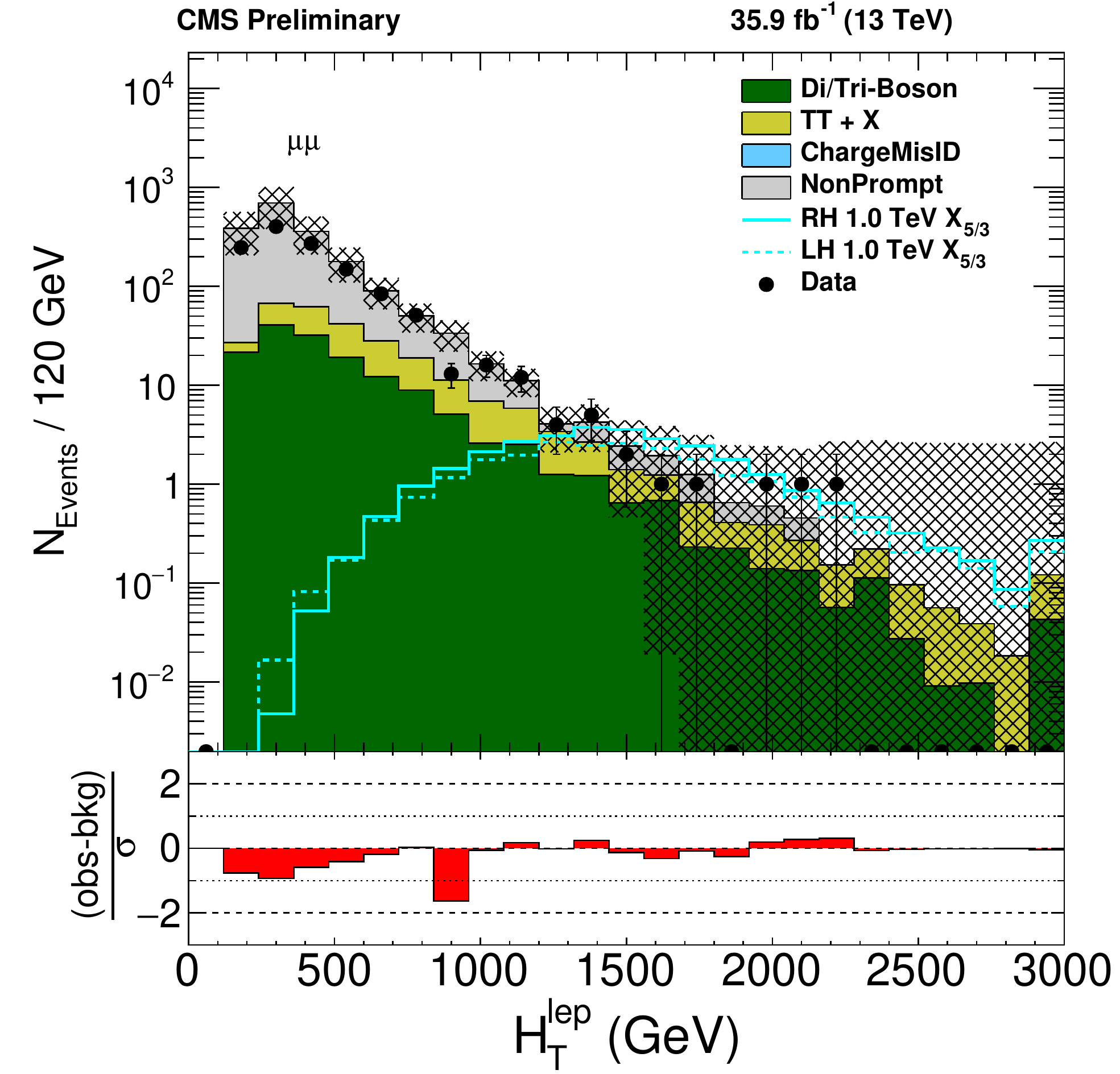}
      \label{XfiveThirdfigb} 
  }\\
  \caption[]{Comparison of data and the SM background expectation after the application of basic selection steps. Shown are the distributions of $H_T^{\textrm{lep}}$ in (a) the $\mu\mu$ and (b) $e\mu$ channel. The figures are taken from~\cite{XfiveThird}.}
  \label{XfiveThirdfig}
\end{figure}

A search for pair production of vector-like quarks, which have an exotic electromagnetic charge of $5/3e$, is performed in the decay channel to a top quark and a $W$ boson using data corresponding to $35.9\;\textrm{fb}^{-1}$~\cite{XfiveThird}.
In order to efficiently suppress SM background processes, the event selection is based on the presence of a lepton pair that features same signs of their electric charges.
Signal processes are furthermore characterized by a rich final state, such that the presence of at least five additional particles, which can be additional leptons or jets, and high values of $H_T^{\textrm{lep}}$ are required for further SM background suppression.
The variable $H_T^{\textrm{lep}}$ is defined as the scalar sum of the transverse momenta of the studied jets and leptons in the events.
The analysis furthermore utilizes the $p_T$ dependent lepton isolation described in Section~\ref{tools}. \\
In Fig.~\ref{XfiveThirdfig} the $H_T^{\textrm{lep}}$ distributions in the $\mu\mu$ and $e\mu$ channel after the application of basic selection steps are presented. 
In neither of the studied categories an excess over the SM expectation is observed after the application of the final event selection.
With the search, right-handed (left-handed) $X_{5/3}$ are excluded below $1.16\;\textrm{TeV}$ ($1.10\;\textrm{TeV}$) at 95\% C.L. 
\section{Search for $\mathbf{W'}$ bosons decaying to a top quark and a bottom quark}
A search for $W'$ bosons decaying into a top quark and a $b$ quark is performed using data corresponding to $35.9\;\textrm{fb}^{-1}$~\cite{B2G17010}. 
In this search, $W'$ boson candidates are reconstructed by building the invariant mass of a jet and a top quark candidate. 
The latter is constructed from a non-isolated lepton, a jet and the measured missing transverse energy.
Event categories are defined based on whether an electron or muon candidate is reconstructed, whether one or two of the leading-$p_T$ jets pass $b$ tag requirements, and whether the event is assigned to the type A or type B event category.
Events are classified as type A or type B depending on the transverse momentum of the top quark candidate and on the transverse momentum of the candidate built from the two $p_T$-leading jets.
In neither of the studied event categories an excess is seen.
Right-handed $W'$ bosons are excluded below $3.4\;\textrm{TeV}$ ($3.6\;\textrm{TeV}$) if the $W'$ boson is assumed to be much heavier (lighter) than the right-handed neutrino postulated in the studied models. 
Furthermore, mass exclusion limits are set for arbitrary coupling strength to left-handed and right-handed fermions. 
\section{Search for $\mathbf{Z'}$ bosons decaying into a top quark pair}
\label{ZPrime}
\begin{figure}[tb]
  \centering
  \subfigure[]{
  \centering
    \includegraphics[width=0.4\textwidth]{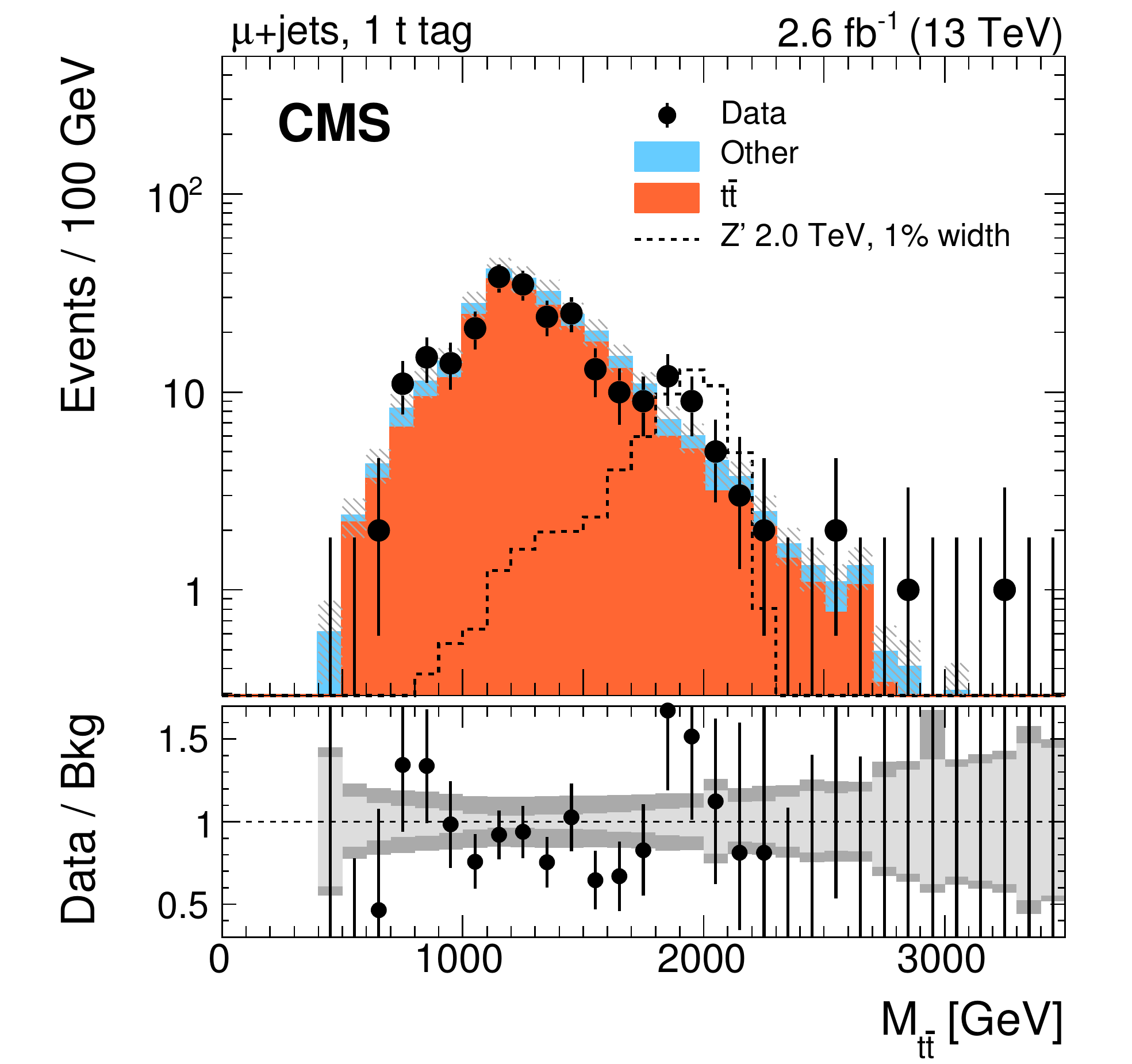}
    \label{ZPrimefiga}
  }
  \subfigure[]{
  \centering
    \includegraphics[width=0.4\textwidth]{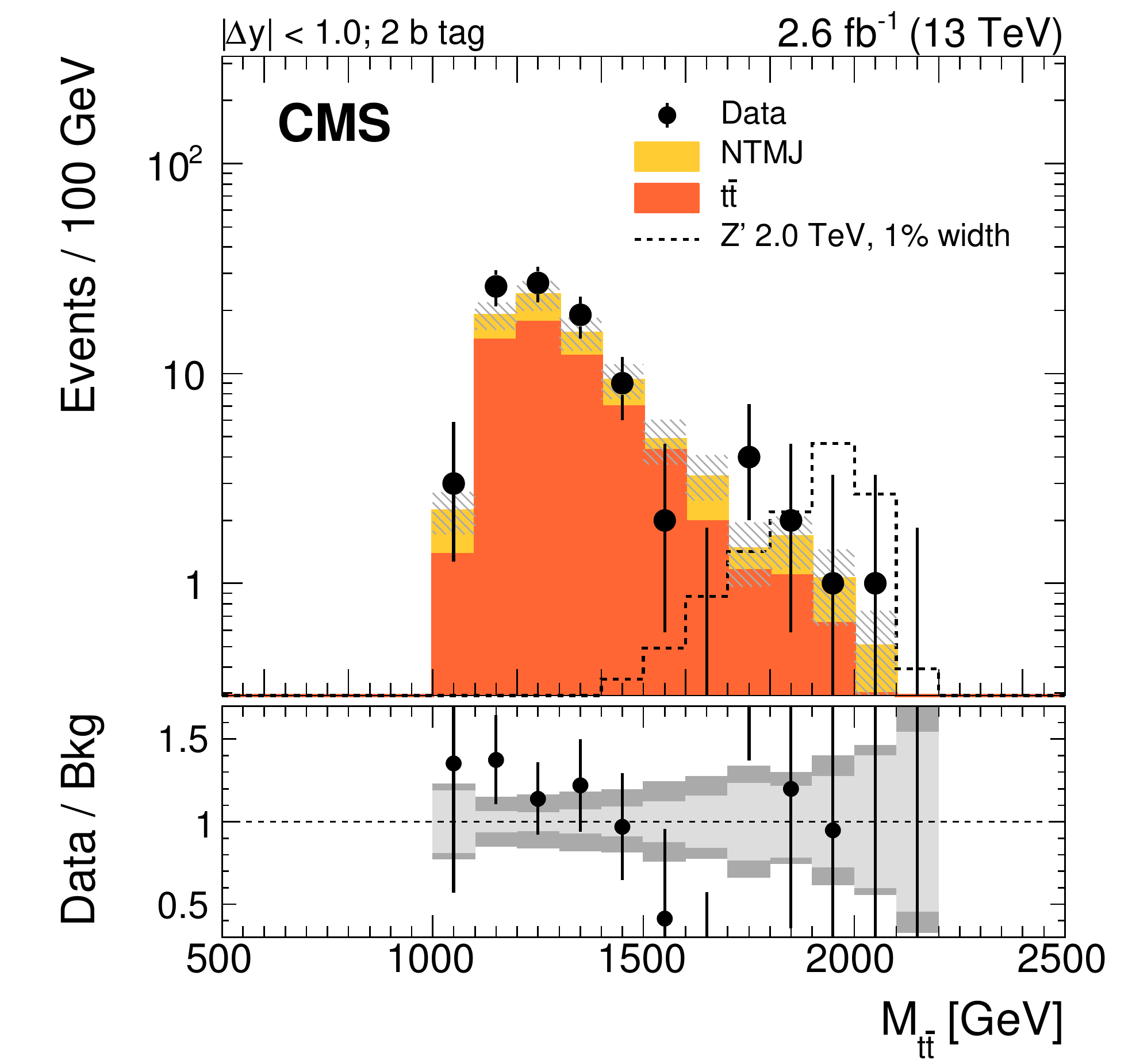}
      \label{ZPrimefigb} 
  }\\
  \caption[]{Comparison of data and the SM background expectation after the application of the signal selection. Shown are the distributions of the invariant mass of the two top quark candidates in (a) the $\mu$+jets channels, where a top-tagged jet is required and (b) the fully hadronic channel, where the top-tagged jets have a rapidity difference $\Delta y$ of $\Delta y <1.0$ and two $b$-tagged subjets are present. The figures are taken from~\cite{ZPrimecomb}.}
  \label{ZPrimefig}
\end{figure}
This section presents a search for $Z'$ bosons decaying into a top quark pair~\cite{ZPrimecomb}. 
The described analysis comprises results of the lepton+jets and the fully hadronic channel.
$2.6\;\textrm{fb}^{-1}$ of $pp$-collision data are analyzed.
The lepton+jets channel is based on the presence of a non-isolated muon or electron candidate. 
Events are categorized based on the number of top-tagged and $b$-tagged jets. 
In total, three categories are defined each for the muon and the electron channel.
The fully hadronic channel contains events with two top-tagged jets. 
The selected events are sorted into six categories depending on the rapidity difference of the top-tagged jets and the number of jets with a $b$-tagged subjet. 
In both the lepton+jets and the fully hadronic analysis, top quark candidates are reconstructed and the invariant mass of the two top quark candidates is built.
In Fig.~\ref{ZPrimefig} two of these invariant mass distributions are presented.
In all studied distributions, a signal would appear as a peak over the falling SM background expectation.
In neither of the studied categories an excess of data over the SM expectation is observed.
With the combined analysis, $Z'$ bosons with a width of 30\% (10\%) are excluded between $500\;\textrm{GeV}$ and $4\;\textrm{TeV}$ ($3.9\;\textrm{TeV}$).
Randall-Sundrum Kaluza-Klein gluons are excluded between $500\;\textrm{GeV}$ and $3.3\;\textrm{TeV}$.
\section{Search for $\mathbf{Z'}$ bosons decaying to a vector-like $\mathbf{T'}$ quark and a top quark}
\begin{figure}[tb]
  \centering
  \subfigure[]{
  \centering
    \includegraphics[width=0.4\textwidth]{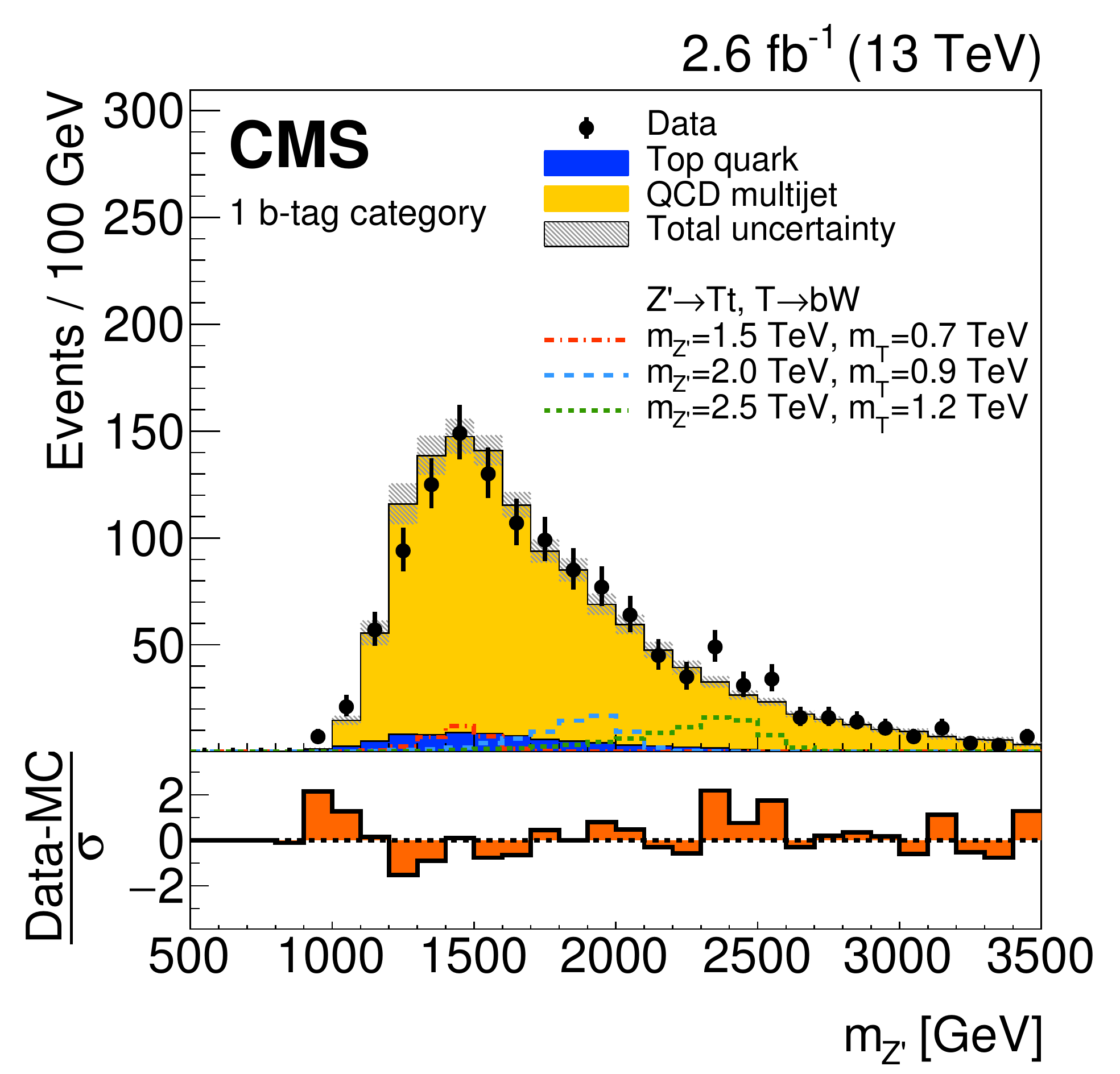}
    \label{ZPrimeTPrimefiga}
  }
  \subfigure[]{
  \centering
    \includegraphics[width=0.4\textwidth]{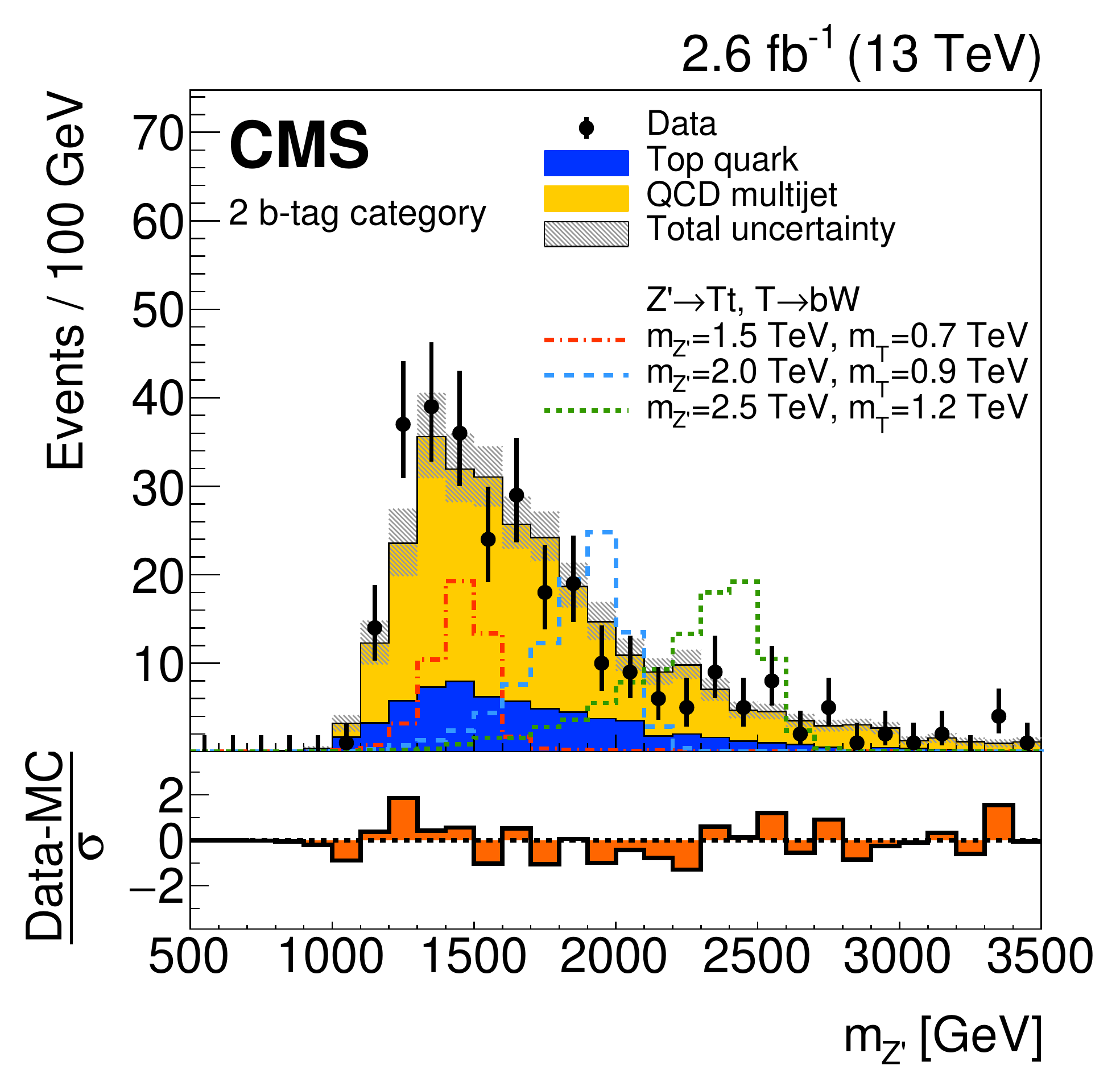}
      \label{ZPrimeTPrimefigb} 
  }\\
  \caption[]{Comparison of data and the SM background expectation after the application of the signal selection. Presented are the distributions of the invariant mass of the reconstructed $Z'$ boson candidate in the category comprising events that contain (a) only the $b$ tagged small-cone jet and (b) a $b$-tagged subjet in addition to the $b$-tagged small-cone jet. The figures are taken from~\cite{ZPrimeTPrime}.}
  \label{ZPrimeTPrimefig}
\end{figure}

Vector-like T' quarks and $Z'$ bosons often appear in the same models. 
In a certain mass range ($m_t +m_{T'} < m_{Z'} < 2m_{T'}$) the decay of a $Z'$ boson into a top quark and a vector-like T' quark might be dominant compared to the decay mode into a top quark pair. 
Since traditional searches, like the search described in Section~\ref{ZPrime}, only cover the decay into a top quark pair, dedicated searches for the decay into a vector-like $T'$ quark and a top quark are developed.
Here, a search in the fully hadronic channel optimized for the $T'$ decay mode into a $b$ quark and a $W$ boson is presented~\cite{ZPrimeTPrime}. 
For the analysis, $2.6\;\textrm{fb}^{-1}$ of data are analyzed.
Events are selected if they contain a boosted $W$ boson candidate from the vector-like $T'$quark decay, a boosted top quark candidate from the $Z'$ boson decay, and a $b$ quark jet candidate by applying top tagging, $W$ tagging and $b$ tagging techniques.
From the tagged jets, a $Z'$ boson candidate is reconstructed.
Selected events are categorized based on whether they contain only the $b$ tagged small-cone jet or also an additional $b$-tagged subjet.
The distributions of the invariant mass of the reconstructed $Z'$ boson candidates in the two event categories are presented in Fig.~\ref{ZPrimeTPrimefig}.
Again, no excess of the data over the SM background expectation is observed.
Upper cross section limits are set for multiple mass combinations of the $Z'$ boson and the vector-like $T'$ quark.
Scans of the branching ratios of the vector-like $T'$ quark into $bW$, $tZ$ and $tH$ are performed.
Furthermore, two benchmark models are tested.
\section{Pair production of leptoquarks decaying to a bottom quark and a tau lepton}
\label{LQbtau}
The last of the presented searches is an analysis~\cite{LQsearch} looking for pair production of new bosons that couple simultaneously to a bottom quark and a tau lepton. They are called leptoquarks.
Data corresponding to an integrated luminosity of $12.9\;\textrm{fb}^{-1}$ is analyzed.
The applied event selection is based on the requirement of the presence of at least one hadronically decaying tau lepton candidate $\tau_h$ and one electron or muon candidate produced in the leptonic decay of the second tau lepton emerging from the leptoquark decay. 
Additionally, each selected event has to contain at least two jets and one of the selected jets has to pass $b$ tagging requirements.
For the final evaluation of the search result a variable referred to as $S_T$ is used. 
The variable $S_T$ is defined as the scalar sum of the transverse momenta of the selected electron or muon candidate, the two jets, the tau lepton candidate, and the measured missing transverse energy. 
While SM background processes tend to populate the lower $S_T$ regions, signal events are characterized by high-$p_T$ objects and thus by events with high values of $S_T$.
Also in this search, the SM prediction describes the data within the statistical and systematic uncertainties. 
No excess is seen.
With the presented search leptoquarks decaying with a branching ratio of 100\% into a bottom quark and a tau lepton are excluded at 95\% C.L. up to masses of 850 GeV.    
\section{Summary}
\begin{justify}
This article presented a few of the most recent analyses of the rich CMS search program for heavy BSM particles decaying into third generation SM quarks. In the presented searches, novel decay channels as well as benchmark models are tested. Most of these searches rely on sub\-struc\-ture tools to maintain sensitivity as the mass scales to be probed increase.\\ 
In this article, only a small subset of available results could be presented. 
Additionally, besides the already exciting results, many more results are about to come with the 2016 data.
All these analyses will help to shed light on the question whether BSM physics exists at the \textrm{TeV} scale or not.
\end{justify}

\end{document}